\documentclass[aps,prc,twocolumn,showpacs,amsmath,amssymb,superscriptaddress]{revtex4-1}

\usepackage{graphicx}
\usepackage{dcolumn}
\usepackage{bm}
\usepackage{hyperref}
\usepackage{comment}

\begin{document}
\preprint{APS/PhysRevC}
\title{Effect of direct reaction channels on deep sub-barrier fusion in
asymmetric systems}

\author{Md. Moin Shaikh}
\affiliation{Nuclear Physics Group, Inter University Accelerator Centre,
Aruna Asaf Ali Marg, Post Box 10502, New Delhi 110067, India}
\author{S. Nath}
\email{subir@iuac.res.in}
\affiliation{Nuclear Physics Group, Inter University Accelerator Centre,
Aruna Asaf Ali Marg, Post Box 10502, New Delhi 110067, India}
\author{J. Gehlot}
\affiliation{Nuclear Physics Group, Inter University Accelerator Centre,
Aruna Asaf Ali Marg, Post Box 10502, New Delhi 110067, India}
\author{Tathagata Banerjee}
\affiliation{Nuclear Physics Group, Inter University Accelerator Centre,
Aruna Asaf Ali Marg, Post Box 10502, New Delhi 110067, India}
\author{Ish Mukul}
\altaffiliation{Presently at TRIUMF, 4004 Wesbrook Mall, Vancouver, British
Columbia, V6T 2A3, Canada}
\affiliation{Nuclear Physics Group, Inter University Accelerator Centre,
Aruna Asaf Ali Marg, Post Box 10502, New Delhi 110067, India}
\author{R. Dubey}
\altaffiliation{Presently at iThemba LABS, National Research Foundation,
PO Box 722, 7129 Somerset West, South Africa}
\affiliation{Nuclear Physics Group, Inter University Accelerator Centre,
Aruna Asaf Ali Marg, Post Box 10502, New Delhi 110067, India}
\author{A. Shamlath}
\affiliation{Department of Physics, School of Mathematical and Physical
Sciences, Central University of Kerala, Kasaragod 671314, India}
\author{P. V. Laveen}
\affiliation{Department of Physics, School of Mathematical and Physical
Sciences, Central University of Kerala, Kasaragod 671314, India}
\author{M. Shareef}
\affiliation{Department of Physics, School of Mathematical and Physical
Sciences, Central University of Kerala, Kasaragod 671314, India}
\author{A. Jhingan}
\affiliation{Nuclear Physics Group, Inter University Accelerator Centre,
Aruna Asaf Ali Marg, Post Box 10502, New Delhi 110067, India}
\author{N. Madhavan}
\affiliation{Nuclear Physics Group, Inter University Accelerator Centre,
Aruna Asaf Ali Marg, Post Box 10502, New Delhi 110067, India}
\author{Tapan Rajbongshi}
\altaffiliation{Presently at Department of Physics, Handique Girls' College,
Guwahati 781001, Assam, India}
\affiliation{Department of Physics, Gauhati University, Guwahati 781014,
India}
\author{P. Jisha}
\affiliation{Department of Physics, University of Calicut, Calicut 673635,
India}
\author{G. Naga Jyothi}
\affiliation{Department of Nuclear Physics, Andhra University,
Visakhapatnam 530003, India}
\author{A. Tejaswi}
\affiliation{Department of Nuclear Physics, Andhra University,
Visakhapatnam 530003, India}
\author{Rudra N. Sahoo}
\affiliation{Department of Physics, Indian Institute of Technology Ropar,
Rupnagar 140001, Punjab, India}
\author{Anjali Rani}
\affiliation{Department of Physics and Astrophysics, University of Delhi,
Delhi 110007, India}

\date{\today}

\begin{abstract}
A steeper fall of fusion excitation function, compared to the
predictions of coupled-channels models, at energies below the lowest barrier
between the reaction partners, is termed as deep sub-barrier fusion hindrance.
This phenomenon has been observed in many symmetric and nearly-symmetric
systems. Different physical origins of the
hindrance have been proposed. This work aims to study the probable effects of
direct reactions on deep sub-barrier fusion cross sections.
Fusion (evaporation residue) cross sections have been measured for
the system $^{19}$F+$^{181}$Ta, from above the barrier down to the energies
where fusion hindrance is expected to come into play.
Coupled-channels calculation with standard Woods-Saxon potential
gives a fair description of the fusion excitation function down to energies
$\simeq 14\%$ below the barrier for the present system. This is in contrast with
the observation of increasing fusion hindrance in asymmetric reactions induced
by increasingly heavier projectiles, \textit{viz.} $^{6,7}$Li, $^{11}$B,
$^{12}$C and $^{16}$O.
The asymmetric reactions, which have not shown any signature
of fusion hindrance within the measured energy range, are found to be induced by
projectiles with lower $\alpha$ break-up threshold, compared to the
reactions which have shown signatures of fusion hindrance. In
addition, most of the $Q$-values for light particles
pick-up channels are negative for the reactions which have exhibited strong
signatures of fusion hindrance, \textit{viz.} $^{12}$C+$^{198}$Pt and
$^{16}$O+$^{204,208}$Pb. Thus, break-up of projectile and particle transfer channels
with positive $Q$-values seem to compensate for the hindrance in fusion deep
below the barrier. Inclusion of break-up and transfer channels within the framework of
coupled-channels calculation would be of interest.
\end{abstract}
\maketitle

Fusion between two nuclei at near barrier energies have been studied quite
extensively in last few decades \cite{dasgupta98,hagino12,back14,canto15}.
Fusion cross sections ($\sigma_{\textrm{fus}}$) have been found to be enhanced
at sub-barrier energies in comparison with the predictions of one-dimensional
barrier penetration model (1D-BPM).
Coupling between the relative motion in the entrance channel, the intrinsic
degrees of freedom of the participating nuclei and nucleon transfer channels
has been invoked to explain the measured fusion excitation functions.
Extending the measurements towards lower energies has immense astrophysical
significance as accurate knowledge of the reaction rates at very low energies
might aid in answering some of the open questions in big bang nucleosynthesis.

Jiang \textit{et al.} \cite{Jiang02} had first observed a steeper fall of
$\sigma_{\textrm{fus}}$ at deep sub-barrier energies in the reaction
$^{60}$Ni+$^{89}$Y, which could not be explained by the coupled-channels (CC)
calculation with standard potential parameters. Subsequently, similar
observations have been reported for many other symmetric and nearly-symmetric
light \cite{Jiang07,Jiang09}, 
medium-light \cite{Jiang08, Stefanini08, Stefanini09, Jiang10, Jiang10a, 
Montanoli10prc, Montanoli10,  Montanoli12, Montanoli13, Jiang14} and
medium-heavy
\cite{Jiang04, Jiang05, Jiang06, Stefanini10, Stefanini15}
systems with some exceptions
\cite{Stefanini06, Stefanini07, Stefanini08, Montanoli10, Montanoli13, Stefanini15,Stefanini17}. 
Two different representations of measured $\sigma_{\textrm{fus}}$ have been
proposed to conclude about fusion hindrance without recourse to model
predictions: (a) the logarithmic derivative of the energy-weighted cross section
\cite{Jiang02},
\textit{viz.},
\begin{equation}
L(E_{\textrm{c.m.}}) = \frac{d[ln(E_{\textrm{c.m.}}\sigma)]}{dE_{\textrm{c.m.}}} = \frac{1}{E_{\textrm{c.m.}}\sigma} \frac{d(E_{\textrm{c.m.}}\sigma)}{dE_{\textrm{c.m.}}}
\end{equation}
and (b) the astrophysical $S$-factor \cite{Jiang04}, \textit{viz.},
\begin{equation}
S(E_{\textrm{c.m.}}) = E_{\textrm{c.m.}}\sigma (E_{\textrm{c.m.}}) \exp(2\pi\eta) .
\end{equation}
Here $E_{\textrm{c.m.}}$ and $\eta$ are the energy available in the centre of
mass (c.m.) frame of reference and the Sommerfeld parameter, respectively.
Observation of continuous increase of $L(E_{\textrm{c.m.}})$ with decreasing
$E_{\textrm{c.m.}}$ and a maximum of $S(E_{\textrm{c.m.}})$ is considered the
clearest signatures of deep sub-barrier fusion hindrance.

A threshold energy has been worked out from systematics \cite{Jiang04a}, below which
fusion hindrance is expected to be observed:
\begin{equation}
\label{Eq_Ethreshold}
E_{\textrm{s}} (\zeta) =
 {0.356 ~\zeta} ^{\frac{2}{3}}
~~~ \textrm{MeV};
\end{equation}
\begin{equation}
L_{\textrm{s}} (\zeta) = 2.33    ~~~\textrm{MeV}^{-1} ,
\end{equation}
where $\zeta = Z_{\textrm{p}}Z_{\textrm{t}}
\sqrt{\frac{A_{\textrm{p}}A_{\textrm{t}}}{A_{\textrm{p}+A_{\textrm{t}}}}}$ is
a parameter characterizing the size of the colliding system; $Z_{\textrm{p}}$
($Z_{\textrm{t}}$) and $A_{\textrm{p}}$ ($A_{\textrm{t}}$) being the atomic and
the mass number of the projectile (target), respectively.

A host of different theoretical approaches has been attempted to explain the
observed change of slope of the fusion excitation functions at energies deep
below the barrier. Mi\c{s}icu and Esbensen \cite{Misicu1,Misicu2} constructed an
ion-ion
potential by double-folding method and added a repulsive core arising from
nuclear incompressibility. Contrary to this \textit{sudden} approach,
Ichikawa \textrm{et al.} \cite{Ichikawa07,Ichikawa09,Ichikawa15} proposed a
smooth transition from sudden to
\textit{adiabatic} potential, as the densities of the participating nuclei
begin to overlap, by imposing a damping factor on the coupling strength.
Dasgupta \textit{et al.} \cite{dasgupta07} suggested that the CC formalism
may itself be inadequate at very low energies and a gradual onset of
quantum decoherence need to be considered. A concise review of all the
theoretical investigations on deep sub-barrier fusion hindrance can be found
in Ref. \cite{back14}. Ichikawa and Matsuyanagi \cite{Ichikawa2015} has also
argued that damping of quantum vibration in the reaction partners near the
touching point is a universal mechanism which is causing hindrance to fusion
deep below the barrier. Pauli exclusion principle has recently been included in
the computation of the bare potential in a new microscopic approach called the
density-constrained frozen Hartree-Fock method \cite{Simenel2017}. Pauli
repulsion has been shown to reduce the tunneling probability, thus, offering
partial explanation of observed hindrance to fusion.

Two different experimental techniques have been adopted to measure
$\sigma_{\textrm{fus}}$ at sub-$\mu$b levels. In the online technique, fusion
products, \textit{i.e.} the evaporation residues (ERs), are separated by an
electromagnetic recoil separator and detected at a background-free site, usually
the focal plane of the separator \cite{Jiang05}. This technique, though direct
and elegant, demands higher recoil energies of the ERs for their efficient
detection. Consequently, fusion reactions between relatively lighter projectiles
on heavier targets, in which ERs do not possess sufficient recoil energy, are
difficult to study experimentally. The difficulty may be overcome in the
off-beam characteristic $\gamma$-ray counting technique \cite{Lemasson}.
Measurements of $\sigma_{\textrm{fus}}$ deep below the barrier have, so far,
been reported for only a handful of asymmetric reactions primarily because of
challenging experimental conditions.

Results from asymmetric reactions seem to suggest that fusion hindrance becomes
increasingly significant with increasing mass and charge of the projectiles. The
reactions $^{6,7}$Li+$^{198}$Pt \cite{Shrivastava09, Shrivastava16} showed no
signs of fusion hindrance. Weak signature of fusion hindrance has recently been
reported for the reaction $^{11}$B+$^{197}$Au \cite{Shrivastava17}. The reaction 
$^{12}$C+$^{198}$Pt \cite{Shrivastava16} exhibited clear sign of fusion
hindrance. Significant hindrance to fusion had been reported in case of
$^{16}$O+$^{204,208}$Pb \cite{dasgupta07}. In all of these reactions measurement
has been extended close to or below the threshold energy (see
Table \ref{q-value}), given by Eq. \ref{Eq_Ethreshold}. One would, thus, expect
to observe fusion hindrance in reactions between projectiles heavier than
 $^{16}$O and heavy targets, below the threshold energy.

To investigate further the role of projectile mass and charge in fusion
hindrance in asymmetric reactions, we extended the measurement of ER cross
sections ($\sigma_{\textrm{ER}}$) for the system $^{19}$F+$^{181}$Ta below the
threshold energy. Measurements of $\sigma_{\textrm{ER}}$ and
$\sigma_{\textrm{fiss}}$ for this reaction had been reported earlier
\cite{Leigh82,Hinde1982,Charity86,Caraley}. $\sigma_{\textrm{fus}}$, which is a
sum of $\sigma_{\textrm{ER}}$ and $\sigma_{\textrm{fiss}}$ was reported by
Nasirov \textit{et al.} \cite{Nasirov2010} in the range of
$E_{\textrm{lab}} = $ 80.0 \textendash 120.0 MeV. At $E_{\textrm{lab}} = 80$
MeV, $\sigma_{\textrm{fiss}}$ is a negligible fraction of
$\sigma_{\textrm{fus}}$. Hence, the $\sigma_{\textrm{ER}}$, reported in this
work, can be taken as $\sigma_{\textrm{fus}}$ for $E_{\textrm{lab}} < 80$ MeV.

The experiment was carried out in two runs at the 15UD Pelletron accelerator
facility of IUAC, New Delhi. A pulsed $^{19}$F beam, with pulse separation of 4
$\mu$s, was bombarded onto a 170 $\mu$g/cm$^2$ thick $^{181}$Ta target on a 20
$\mu$g/cm$^2$ $^{\textrm{nat}}$C backing. Measurements were performed at beam
energies ($E_{\textrm{lab}}$) in the range of 73.7 \textendash 123.8 MeV using
the Heavy Ion Reaction Analyzer (HIRA) \cite{Sinha94}. ERs formed in complete
fusion of the reaction partners were separated from orders-of-magnitude larger
background events by the HIRA. Two monitor detectors were placed at laboratory
angle ($\theta_{\textrm{lab}}$) $15.5^\circ$ with respect to beam direction, in
the horizontal plane, inside the target chamber for absolute normalization of
$\sigma_{\textrm{ER}}$. A thin (30 $\mu$g/cm$^2$)
$^{\textrm{nat}}$C foil was placed 10 cm downstream from the target to reset
ER charge states to equilibrium distribution. A multi-wire proportional counter
(MWPC), having an active area of 15.0$\times$5.0 cm$^{2}$, was used to detect
ERs at the focal plane of the HIRA. A very thin (0.5 $\mu$m) mylar foil was used
as the entrance window of the MWPC, filled with isobutane at 3 mbar pressure, to
minimize loss of energy for ERs. The HIRA was operated at 0$^{\circ}$ with 10
msr acceptance. Time interval between the arrival of a particle at the focal
plane and the beam pulse was recorded as a measure of ER time of flight (TOF).
Yield of ERs were extracted from the coincidence spectrum between energy loss 
($\Delta E$), obtained from the cathode of MWPC, and TOF.

\begin{figure}
\resizebox{\columnwidth}{!}{\includegraphics[angle=0]{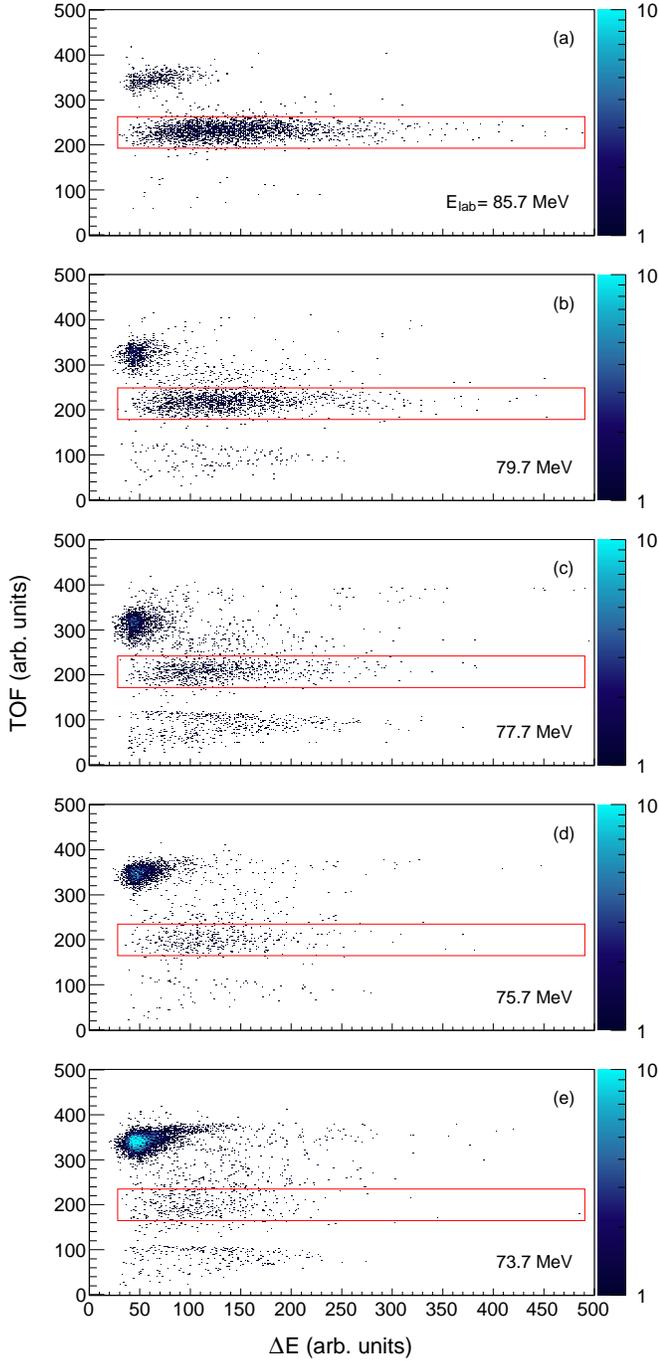}}
\caption{\label{spec}(Color online) Scatter plots between $\Delta E$ and TOF
of the events recorded at the focal plane of the HIRA for $^{19}$F+$^{181}$Ta,
from (a) $E_{\textrm{lab}} = 85.7$ MeV
$\left(\frac{E_{\textrm{c.m.}}}{V_{\textrm{B}}} \simeq 1.01\right)$ to
(e) $E_{\textrm{lab}} = 73.7$ MeV
$\left(\frac{E_{\textrm{c.m.}}}{V_{\textrm{B}}} \simeq 0.86\right)$.
$V_{\textrm{B}}$ is the Coulomb barrier in c.m. frame of reference. ERs are
enclosed within a rectangular gate in each panel. See text for details.}
\end{figure}

The most challenging aspect of measuring $\sigma_{\textrm{fus}}$ in this
experiment was to identify the ERs unambiguously. It is clearly observed from
Fig. \ref{spec} that separating the ERs from the scattered projectile-like
particles becomes increasingly challenging with decreasing $E_{\textrm{lab}}$.
Near the barrier ($E_{\textrm{lab}} = 85.7$ MeV), the ERs are clearly
identified (Fig. 1.(a)). At $E_{\textrm{lab}} = 73.7$ MeV, ERs started merging
with the background (Fig. 1.(e)), thus setting the lower energy limit in the
present measurement. The other major challenge was to estimate the
transmission efficiency of the HIRA ($\epsilon_{\textrm{HIRA}}$) accurately,
which is crucial for extraction of absolute $\sigma_{\textrm{ER}}$ from data.
$\epsilon_{\textrm{HIRA}}$ was calculated using the semi-microscopic Monte
Carlo code \textsc{ters} \cite{Nath08}. Details of the data analysis
method can be found in Refs. \cite{Nath10,Rajbongshi2015}.
 
\begin{figure}
\resizebox{\columnwidth}{!}{\includegraphics[angle=0]{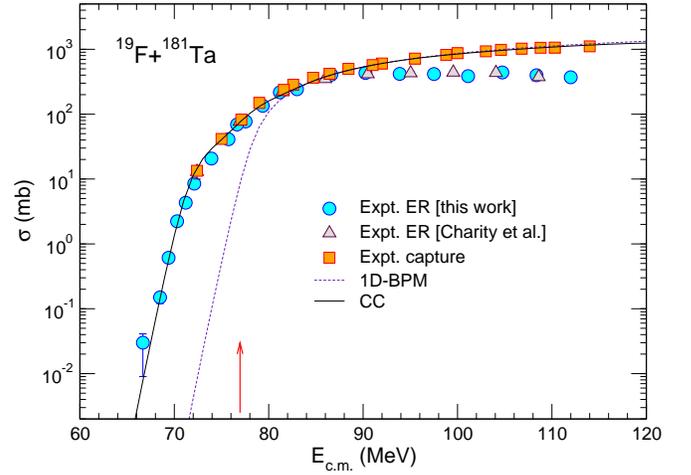}}
\caption{\label{fusion}(Color online) Measured $\sigma_{\textrm{ER}}$ and
$\sigma_{\textrm{fus}}$, as a function of $E _{\textrm{c.m.}}$, for the system 
$^{19}$F + $^{181}$Ta. Data points shown with solid triangles had been reported
by Charity \textit{et al.} \cite{Charity86} whereas experimental 
$\sigma_{\textrm{fus}}$ were obtained from Ref. \cite{Nasirov2010}. Results from
CC model calculation are also shown. The vertical arrow indicates position of
the Coulomb barrier.}
\end{figure}

The previously measured $\sigma_{\textrm{fus}}$ and $\sigma_{\textrm{ER}}$,
along with data from the present investigation are presented in Fig.
\ref{fusion}. One may note that data from this work match with data from the
earlier measurement within the experimental uncertainties in the overlapping
energy region. The experimental fusion excitation function has also been
compared with the predictions of the coupled-channels code \textsc{ccfull}
\cite{ccfull} in Fig. \ref{fusion}. In the calculation, Woods-Saxon potential
parameters, \textit{viz.} depth (\texttt{V$_{0}$}), radius (\texttt{r$_{0}$})
and diffuseness (\texttt{a}), were chosen as 104.5 MeV, 1.12 fm and 0.70 fm,
respectively, so as to reproduce the excitation function above the barrier.
The output with no coupling configuration has been taken as the
1D-BPM cross sections. As both the target and the
projectile are odd-even nuclei, the deformation parameters have been taken as
the average of their neighboring even-even nuclei. It is observed from Fig.
\ref{fusion} that couplings have very insignificant effect at energies above the
barrier and both 1D-BPM and CC predictions match with the
experimental $\sigma_{\textrm{fus}}$. As energy decreases the 1D-BPM prediction
starts to underestimate the experimental data. The coupling to the inelastic
excitations of target and projectile describes the data quite satisfactorily
up to the lowest energy reported in this work. Thus, no sign of fusion hindrance
compared to CC prediction has been observed in the measured energy range.

\begin{figure}[ht]
\resizebox{\columnwidth}{!}
{\includegraphics[trim=0 0 0 0, clip=true, angle=0]{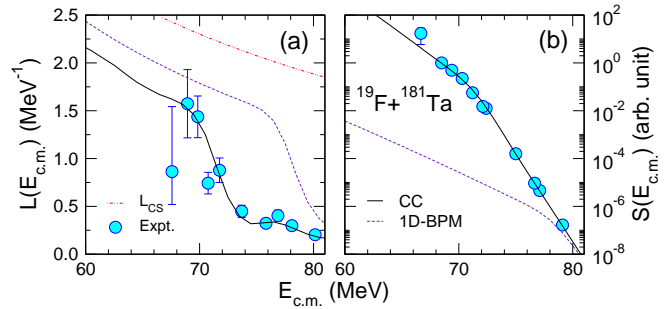}}
\caption{\label{LS}(Color online) (a) The logarithmic derivative of the
energy-weighted cross section and (b) the astrophysical $S$-factor, as a
function of $E _{\textrm{c.m.}}$, for the system $^{19}$F + $^{181}$Ta. Results
from CC model calculation are also shown. The double-dot-dashed line in panel
(a) represents the logarithmic derivative for a constant $S$-factor,
$L_{\textrm{CS}}(E_{\textrm{c.m.}})$.}
\end{figure}

To amplify any possible discrepancy between the experimental and theoretical
fusion excitation functions in the measured energy range, $L(E_{\textrm{c.m.}})$
has been plotted in Fig. \ref{LS}(a). It is noted that the CC predictions
explain the data quite satisfactorily throughout the measured energy range. No
pronounced change of slope in the deep sub-barrier energy region is observed.
The double-dot-dashed line represents the logarithmic derivative for a constant
$S$-factor,
$L_{\textrm{CS}}(E_{\textrm{c.m.}}) = \frac{\pi\eta}{E_{\textrm{c.m.}}}$
\cite{Jiang04}. We observe that $L_{\textrm{CS}}(E_{\textrm{c.m.}})$ curve
lies much above the data and the CC results. There is no cross-over between
$L_{\textrm{CS}}(E_{\textrm{c.m.}})$ and $L(E_{\textrm{c.m.}})$, which is
observed in systems exhibiting fusion hindrance \cite{Jiang10, Shrivastava16}.
The intersection between these two curves is taken as the experimental threshold
for fusion hindrance. Therefore, no threshold of fusion hindrance is observed
for the reaction $^{19}$F+$^{181}$Ta till $\simeq 12$ MeV below the barrier
(equivalent $\simeq 0.86V_{\textrm{B}}$). $S(E_{\textrm{c.m.}})$ has been
plotted in Fig. \ref{LS}(b). Neither the maximum nor any saturation of the
$S$-factor is observed within the measured energy range.

We next compare the observations for the system $^{19}$F+$^{181}$Ta with results
of other asymmetric systems mentioned earlier. The present reaction does not
follow the systematic observations of increasingly stronger fusion hindrance
with increasing mass and charge of the projectile. The contrast between the
systems $^{16}$O+$^{204, 208}$Pb and $^{19}$F+$^{181}$Ta is particularly
striking. Hindrance to fusion starts showing up from $\simeq 5$ MeV below the
barrier (equivalently $\simeq 0.94V_{\textrm{B}}$) for the  $^{16}$O-induced
reaction, whereas no hindrance is observed even $\simeq 12$ MeV below the
barrier (equivalently $\simeq 0.86V_{\textrm{B}}$) for the $^{19}$F-induced
reaction.

Recent theoretical investigations \cite{Ichikawa2015,Simenel2017} point to the
generic nature of fusion hindrance deep below the barrier and acknowledge
(a) gradual onset of decoherence, (b) transition from nucleus-nucleus sudden to
one-nucleus adiabatic potential and (c) Pauli repulsion as the probable
contributors in the phenomenon. In case of non-observation of fusion hindrance
in a particular reaction one must, therefore, look into other factors which are
specific to that reaction.

Break-up of the projectile and presence of light particle transfer channels are
known to affect fusion between two nuclei. We note that the reactions showing
no hindrance to fusion are induced by projectiles with low $\alpha$-particle
break-up threshold: 1.474 MeV, 2.468 MeV and 4.013 MeV for $^{6}$Li, $^{7}$Li
and $^{19}$F, respectively. On the other hand, reactions exhibiting fusion
hindrance are induced by projectiles with much higher $\alpha$-particle
break-up threshold: 8.665 MeV, 7.366 MeV and 7.162 MeV for $^{11}$B, $^{12}$C
and $^{16}$O, respectively. $n$, $2n$, $p$, $2p$, $^{2}$H, $^{3}$H, $^{3}$He
and $^{4}$He pick-up transfer $Q$-values for the asymmetric reactions are
presented in Table \ref{q-value}. One may also note in Table \ref{q-value} that
many of the particle transfer $Q$-values are positive for the reactions
showing no fusion hindrance, \textit{viz.} $^{6,7}$Li+$^{198}$Pt and
$^{19}$F+$^{181}$Ta. On the contrary, the reactions showing clearest
signatures of fusion hindrance \textit{i.e.} $^{12}$C+$^{198}$Pt and
$^{16}$O+$^{208}$Pb have all but one negative particle transfer $Q$-values.
We may, therefore, conclude that presence of break-up and particle transfer
channels are aiding in sub-barrier fusion. The effects of fusion hindrance,
which is expected to be manifested in all reactions at energies below the
threshold energy, appears to have been \textit{compensated} in the presence of
these direct reaction channels. A comprehensive theoretical investigation is
needed to strengthen and quantify this conclusion.

\begin{table*}
\centering
\caption{Details of the asymmetric systems considered in this work.
$E_{\textrm{min}}$ is the lowest energy in the c.m. frame of
reference at which measurement has been reported. $Q_{\textrm{CN}}$ is the
$Q$-value for compound nucleus formation. $Q_{\textrm{x}}$ is the pick-up
transfer $Q$-values, where x stand for $n$, $2n$, $p$, $2p$, $^{2}$H, $^{3}$H,
$^{3}$He and $^{4}$He.}
\label{q-value} 
\begin{ruledtabular}    
\begin{tabular}{lcccccccccccccc}
System & $Z_{\textrm{p}}Z_{\textrm{t}}$ & $\zeta$ & $V_{\textrm{B}}$ & $E_{\textrm{s}}$ & $E_{\textrm{min}}$ & $Q_{\textrm{CN}}$ & $Q_{n}$ & $Q_{2n}$ & $Q_{p}$ & $Q_{2p}$ & $Q_{^2\textrm{H}}$ & $Q_{^3\textrm{H}}$ & $Q_{^3\textrm{He}}$ & $Q_{^4\textrm{He}}$\\
 & & & (MeV) & (MeV) & (MeV) & (MeV) & (MeV) & (MeV) & (MeV) & (MeV) & (MeV) &
(MeV) & (MeV) & (MeV)\\
\hline
$^{19}$F+$^{181}$Ta & 657 & 2724 & 77.9 & 69.4 & 66.7 & -23.67 & -0.98 & 0.48 & 6.90 & 1.32 & 6.27 & 10.53 & 5.59 & 11.99\\

$^{16}$O+$^{208}$Pb \cite{dasgupta07}& 656 & 2529 & 77.0 & 66.1 & 68.7 & -46.48 & -3.22 &-1.92 &-7.40 &-10.86& -5.11& -1.18 & -5.95& 5.25 \\

$^{16}$O+$^{204}$Pb \cite{dasgupta07}& 656 & 2527 & 77.3 & 66.0 & 68.7 & -44.52 & -4.25 & -3.12& -6.04& -7.82 & -4.74& -1.18  & -3.94 & 6.70  \\

$^{12}$C+$^{198}$Pt \cite{Shrivastava16}& 468 & 1574 & 56.0 & 48.2 & 47.0 & -13.95 & -2.61 &-0.28 &-6.99 &-9.63 & -3.33& 1.69  & -3.25& 7.27 \\

$^{11}$B+$^{197}$Au \cite{Shrivastava17}& 395 & 1275 & 47.4 & 41.9 & 37.9 & 5.00   & -4.70 &-6.47 &10.17 & 3.87 &  7.20& 9.27  & 7.20 & 11.96 \\

$^{7}$Li+$^{198}$Pt \cite{Shrivastava16}& 234 & 608  & 28.5 & 25.6 & 19.3 & 8.82   & -5.52 &-7.31 & 8.33 & 0.86 &  3.09& 4.09  & 2.46 & 8.77\\

$^{6}$Li+$^{198}$Pt \cite{Shrivastava09}& 234 & 565  & 28.9 & 24.3 & 19.6 & 8.53   & -0.30 &-4.12 &-3.32 &-10.46& 8.68 & 4.53  & 1.28 & 4.57\\

\end{tabular}
\end{ruledtabular}
\end{table*}

In summary, we have measured $\sigma_{\textrm{ER}}$ for the reaction
$^{19}$F+$^{181}$Ta down to $\simeq 14\%$ below the barrier. As
$\sigma_{\textrm{fiss}}$ had earlier been shown to be insignificant below the
barrier, $\sigma_{\textrm{ER}}$ measured in this work has been considered as
$\sigma_{\textrm{fus}}$ at $E_{\textrm{lab}} < 80$ MeV. CC calculation with
standard Woods-Saxon potential has reproduced $\sigma_{\textrm{fus}}$ quite
satisfactorily. Thus, fusion hindrance has not been observed in this reaction,
though measurement has been extended below the threshold energy. We have
compared our observation with results from other asymmetric reactions. It has
been found that reactions induced by projectiles with low $\alpha$-particle
break-up threshold and having many light particles transfer channels with
positive $Q$-values did not show fusion hindrance even below the threshold
energy. On the other hand, reactions induced by strongly bound projectiles and
having most of the light particles transfer channels with negative $Q$-values
exhibited fusion hindrance. This is in contradiction with the recent
observation of increasingly stronger fusion hindrance in asymmetric systems
with increasing mass and charge of the projectile. Extending the measurement
deeper below the barrier for $^{19}$F+$^{181}$Ta and measuring
$\sigma_{\textrm{fus}}$ for more asymmetric systems will strengthen our
conclusion. CC calculation including the effects of projectile break-up and
particle transfer to reproduce fusion excitation functions deep below the
barrier would complement the challenging experimental efforts.

The authors thank the Pelletron staff of IUAC for providing beams of excellent
quality throughout the experiment, Mr. Abhilash S. R. for assistance in
fabricating the target and Mr. T. Varughese for support during the experiment.
The authors thank Mr. Rohan Biswas for useful discussions.

\end{document}